\documentclass[aps,prl,reprint,groupedaddress]{revtex4-1}

\usepackage{amsmath}
\usepackage{amssymb}
\usepackage{hyperref}
\usepackage{graphicx}
\usepackage{epstopdf}
\usepackage{textcomp}
\usepackage{soul}
\usepackage{color}

\begin{document}


\title{Structure and Dynamics of the Instantaneous Water/Vapor Interface Revisited \\ by Path-Integral and Ab-Initio Molecular Dynamics Simulations}


\author{Jan Kessler}
\author{Hossam Elgabarty}
\affiliation{Institute of Physical Chemistry and Center of Computational Sciences, Johannes Gutenberg University Mainz, Staudinger Weg 9, D-55128 Mainz, Germany}
\author{Thomas Spura}
\author{Kristof Karhan}
\author{Pouya Partovi-Azar}
\affiliation{Department of Chemistry, University of Paderborn, Warburger Str. 100, D-33098 Paderborn, Germany}
\author{Ali A. Hassanali}
\affiliation{The Abdus Salam International Centre for Theoretical Physics, Strada Costiera 11, I-34151 Trieste, Italy}
\author{Thomas D. K\"uhne}
\email{tdkuehne@mail.upb.de}
\affiliation{Department of Chemistry and Paderborn Center for Parallel Computing, University of Paderborn, Warburger Str. 100, D-33098 Paderborn, Germany}
\affiliation{Institute for Lightweight Design with Hybrid Systems, Warburger Str. 100, D-33098 Paderborn, Germany}


\date{\today}

\begin{abstract}
The structure and dynamics of the water/vapour interface is revisited by means of path-integral and second-generation Car-Parrinello \textit{ab-initio} molecular dynamics simulations in conjunction with an instantaneous surface definition [A. P. Willard and D. Chandler, J. Phys. Chem. B {\bf 114}, 1954 (2010)]. In agreement with previous studies, we find that one of the OH bonds of the water molecules in the topmost layer is pointing out of the water into the vapor phase, while the orientation of the underlying layer is reversed. Therebetween, an additional water layer is detected, where the molecules are aligned parallel to the instantaneous water surface. 
\end{abstract}


\maketitle

\section{Introduction}

The water/vapor interface is ubiquitous in many interesting applications involving both chemical processes and phenomena in biology and
aqueous chemistry\cite{Ball2008}.  A better understanding of the interfacial behavior of water is crucial for studying phenomena as diverse as protein folding \cite{Dobson2003,Dill2012}, the structure and function of biological membranes and membrane proteins \cite{Killian2000}, the Hofmeister effect \cite{Collins1985}, electrochemical processes in aqueous batteries \cite{Luo2010}, and the remarkable organic catalysis on water \cite{Narayan2005,Jung2010,Karhan2014}, to mention just a few. Despite long study, understanding the molecular aspects of hydrophobic solvation continues to be an area of active research \cite{HummerPNAS1,HummerPNAS2,LumChandlerWeeks1999,HummerReview,ChandlerReview2005}.

The simplest case of an aqueous-hydrophobic boundary is the water/vapor interface, where the vacuum can be considered as the ``ultimate hydrophobe''\cite{Marx2004}, with no attractive or repulsive van der Waals forces between the water and the hydrophobic phase. Due to its prototypical character, this particular interface has been studied extensively, both experimentally \cite{Schwartz1990,Du1993,Du1994,BAR2000,Scatena2001,Wilson2002,Gopalakrishnan2005,Gan2006,Cappa2008,Sovago2008,SkinnerJACS2011,BonnJACS2011,SkinnerNature2011,Zhang2011,Torun2014} and theoretically\cite{Rossky1984,Benjamin1994,Dang1997,Morita2000,Morita2002,kuo2004,Mucha2005,Buch2005,Moore2005,Tobias2006,Fan2009,Auer2009,Morita2009,Stirnemann2010,Kuhne2011,Tobias2011,Nihonyanagi2011,SkinnerJCP2011,Morita2012,Nagata2012,Zhang2013,Sulpizi2013,Martyna2015,Zhang2015}. Moreover, there is an ongoing debate in the literature regarding the pH of water at the water/vapor interface. Different experimental and theoretical studies have suggested the possibility that the pH at the surface of water could be either basic or acidic \cite{BuchPNAS,Buch2007,Saykally2008,Beattie2009,Voth2009,ColussiPNAS}. More recently, it has also been proposed that enhanced charge transfer at the interface could rationalize the existence of the negative charge at the surface of water \cite{Roke2011,Jungwirth2012}. In any case, even the bare water/vapor interface remains highly contentious.

The broken symmetry at the water/vapor interface restricts the ability of water molecules to form H-bonds like they do in the bulk. One of the most common quantities used to characterize the average configuration of water molecules is the ratio of water molecules that corresponds to single-donor (SD), double-donor (DD), or non-donor (ND) configurations. Within this framework, perfect crystalline bulk ice consists of 100\% DD configurations, whereas bulk liquid water and its surface are expected to have a mixture of all three configurations with varying ratios induced by thermal fluctuations. Various experimental techniques have been applied in order to probe the structural properties at the interface. In particular, using surface-sensitive vibrational sum-frequency generation (SFG) spectroscopy, Shen and coworkers observed a complete suppression of the free OH peak at 3700~$\textrm{cm}^{-1}$ by titrating the dangling OH groups at the water/vapor interface with methanol \cite{Du1993, Du1994}. From these measurements, they estimated that the vapor/water interface consists of 25\% SD and 75\% DD configurations. This implies that a significant fraction of the water molecules at the interface possess dangling OH bonds that are protruding out of the water phase into the vacuum. 
In later studies, dangling OH groups at the interface have been confirmed by SFG measurements of others \cite{Eisenthal1996,Richmond2002,Gopalakrishnan2005,Gan2006} and molecular dynamics (MD) simulations \cite{Rossky1984,Dang1997,Morita2000,kuo2004,Buch2005,Perry2006,Kuhne2011,Martyna2015}. Nevertheless, other experiments using X-ray absorption \cite{Wilson2002,Cappa2008} and SFG spectroscopy \cite{Scatena2001}, as well as \emph{ab initio} MD (AIMD) simulations\cite{kuo2004} have found a much larger fraction of ``accepter-only'' water configurations. However, quantifying the relative occurrence of the various water configurations entails a multitude of computational challenges, which are due to the large dipole moment, polarizability and nuclear quantum effects (NQE) of liquid water, as well as nonadditive cooperative effects of H-bonds \cite{Stillinger1980}. Hence, the predictive power of such simulations critically depends on the accuracy of the employed interaction potential, but also on the involved time and length scales to minimize statistical uncertainties and single-particle finite-size effects \cite{Kuhne2009}. In case of the water/vapor interface, an even larger system has to be considered in order to stabilize a two-dimensional periodic water slab with a sufficiently large vacuum portion on top of it to eliminate spurious long-range interactions at the surface. 

In an earlier work, we studied the water/vapor interface from {\em first principles} using the second-generation Car-Parrinello AIMD method \cite{Kuhne2011,Kuhne2007,Kuhne2014}. 
In this study, no evidence of a significant occurrence of ND water configurations at the water surface was observed. However, determining the exact location of the water/vapor interface is quite ambiguous due to the inherently large spatial and temporal fluctuations of the instantaneous interface. 
Yet, the particular definition of the water surface can significantly affect the quantitative values of DD, SD and ND water molecules near the water/vapor interface. As a consequence, there have been several attempts to refine the definition of the interface at a molecular level that accounts for the instantaneous fluctuations \cite{Percus1987,Gomes1999,Rossky2001,Tarazona2003}. More recently, an instantaneous interface definition to elucidate the presence of enhanced structural correlations at the interface that are not captured by a criterion based on a time-averaged density profile has been proposed \cite{Willard2010}.

However, since water mainly consists in a large part of light H atoms, NQE, such as the quantum mechanical zero-point energy (ZPE) and tunneling effects, are potentially essential to obtain the correct quantitative, and sometimes even qualitative behavior, and thus should be explicitly taken into account  \cite{MarxScience1997,MarxNature1999,MarxNature2002,Morrone2007,Lin2011,Nagata2012,RPMDrev2013,Ceriotti2013,SpuraFM2015}. Although it is well known that NQE generally weaken intermolecular H-bonding, resulting in a faster rotational and translational dynamics and at the same time less structured liquid, there is an ongoing debate regarding the magnitude of this effect \cite{MillerWater2005,KusalikWater2006,PaesaniWater2006,Habershon2009,SpuraFM2015}. In fact, Habershon et al. have shown that there are competing NQE in bulk water: on the one hand the enhanced quantum fluctuations of the protons result in stronger H-bonding, increased structure and slower dynamics, while on the other hand, the enhanced librational modes weakens the H-bonds, reduces the structure and increases the diffusion \cite{Habershon2009}. Moreover, there exists considerable interest on the impact of NQE on the water/vapor interface. In particular, both experimental and theoretical studies have shed light on the presence of isotope fractionation effects observed between the liquid water and vapor phase, which is due to NQE \cite{Wesolowski1994,Cappa2003,MarklandBerne2012,CeriottiManlopoulos2013,CeriottiMarkland2014}. In general, the role of NQE on the structural and dynamical properties of the water/vapor interface remains poorly understood.

In this work, we revisit the orientation of water molecules at the water/vapor interface, as well as the structure and dynamics of the corresponding H-bond network  based on AIMD and including NQE, by means of path-integral MD (PIMD) simulations. At variance to our previous study \cite{Kuhne2011}, the improved instantaneous surface definition of Willard and Chandler is employed to identify the individual water layers and to investigate the orientational distribution, H-bond network, as well as dynamics of interfacial water.

The remainder of this paper is organized as follows. In Section 2, we summarize the finite temperature path-integral technique to account for ZPE and tunnelling effects, as well as the H-bond and instantaneous surface definitions. Thereafter, in Section 3, we describe the computational details of our AIMD and PIMD simulations. Our results are presented and discussed in Section 4 before concluding the paper in Section 5. 

\section{Computational Methods}

\subsection{Path-Integral Formalism}


In the PIMD method, all quantum particle are replaced by harmonic P-bead ring-polymers, which are then treated classically. This is to say that quantum mechanical properties can be exactly calculated by sampling the path-integral phase space using MD, since the extended ring-polymer system is isomorphic to the original quantum system \cite{Feynman1948,Chandler1981,Parrinello1984}. 
For this purpose, the canonical quantum partition function $Z(\beta)$ is expressed in terms of the inverse temperature $\beta^{-1}=k_{B}T$, i.e. 
\begin{equation}
  \label{eq:1}
  Z(\beta)=\textrm{Tr} \left [ e^{-\beta\hat{H}}\right ] = \textrm{Tr}
  \left [ \left(e^{-\beta_{P}\hat{H}} \right)^{P}  \right ] =\lim_{{P \to \infty}}
  Z_{P}(\beta) ,
\end{equation}
where $\hat{H}=\hat{T}+\hat{V}$ is the Hamilton operator. In other words, the origin of the method is the notion that the finite temperature density matrix $e^{-\beta\hat{H}}$, which corresponds to the square of the wavefunction at low and to the Maxwell-Boltzmann distribution at high temperature, can be decomposed into a product of density matrices, each at higher effective temperature $\beta_{P}=\beta/P$. 
In any case, Eq.~\ref{eq:1} is a direct consequence of the Trotter theorem and implies that in the limit $P \rightarrow \infty$, sampling $Z_P$ classically is equivalent to the exact canonical quantum partition function $Z$ \cite{Parrinello1984}. 

Inserting $P-1$ complete sets of position eigenstates and introducing momenta using the standard Gaussian integral, as well as the symmetric Trotter splitting to decompose $\hat{H}$, 
\begin{equation}
  \label{eq:3}
  Z_{P}=\mathcal{N} \int d^{NP}\,\mathbf{r}\int
  d^{NP}\,\mathbf{p}\,e^{-\beta H_{P}(\{\mathbf{r}\},\{\mathbf{p}\})},
\end{equation}
which can be readily sampled by MD. Herein, $\mathcal{N}^{-1}=(2 \pi \hbar)^{NP}$ is a normalization constant, while $\mathbf{r}$ and $\mathbf{p}$ are the positions and momenta of all $N$ particles. The so-called bead-Hamiltonian $H_{P}(\{\mathbf{r}\},\{\mathbf{p}\})$ that describes the interactions between all $N\times P$ beads, reads as 
\begin{align}
  \label{eq:4}
  H_{P}(\{\mathbf{r}\},\{\mathbf{p}\}) &=\sum_{k=1}^{P} \left [ \sum_{i=1}^{N} \left (
    \frac{(\mathbf{p}_{i}^{(k)})^{2}}{2m_{i}^{(k)}} +
    \frac{m_{i}\omega_{P}^{2}}{2} \left ( \mathbf{r}_{i}^{(k)} -
    \mathbf{r}_{i}^{(k+1)} \right)^{2}  \right )
    \right. \\ &+\frac{1}{P}V( \{ \mathbf{r}_{i}^{(k)} \} )
    \left. \vphantom{\frac12} \right ] _{ \mathbf{r}_{i}^{(P+1)} =
    \mathbf{r}_{i}^{(1)} }, \nonumber
\end{align}
where $P$ is the number of imaginary-time slices, $m_i$ are the particle masses and $\omega_{p}=P/\beta=\beta_P^{-1}$ is the angular frequency of the harmonic spring potential between the beads. As a consequence of the trace of Eq.~\ref{eq:1}, $H_{P}(\{\mathbf{r}\},\{\mathbf{p}\})$ is isomorphic to a classical closed ring-polymer, thus $\mathbf{r}_{i}^{(P+1)} = \mathbf{r}_{i}^{(1)}$ \cite{Chandler1981}.  

For the purpose to reduce the computationally dominating effort to evaluate the intermolecular long-range electrostatic interactions $P$ times, the ring-polymer contraction scheme is employed \cite{Markland2008}. To that extent, the Hamiltonian is split into its inter- and intramolecular contributions, where the former is
limited to a single Ewald sum at the centroid of the closed ring-polymer:
\begin{equation}
  \label{eq:5}
  \mathbf{r}_{i}^{{c}}=\frac{1}{p}\sum_{k=1}^{p}\mathbf{r}_{i}^{(K)}.
\end{equation}


At variance to the original PIMD method, the partially adiabatic centroid MD (ACMD) approach permits to approximately compute even dynamical properties within the path integral scheme \cite{Hone2006}. To that extent, the effective masses of the ring-polymer beads are chosen so as to recover the correct dynamics of the ring-polymer centroids. Specifically, the elements of the Parrinello-Rahman mass-matrix are selected so that the vibrational modes of all ring-polymer beads, except for the centroid, are shifted to a frequency of 
\begin{equation}
  \label{eq:8}
  \Omega=\frac{P^{P/P-1}}{\beta\hbar},  
\end{equation}
which allows for integration time-steps close to the ionic resonance limit \cite{Habershon2008}. 

\subsection{Hydrogen Bond Definition and Kinetics}

In order to determine whether two water molecules are H-bonded or not, different energetic and geometric criteria have been proposed \cite{Stillinger1980, Luzar1993, kuo2004, Buch2008, Ceriotti2014}. 
Following Skinner and coworkers, we use the potential of mean force (PMF) as calculated from the radial-angle joint distribution function to define a H-bond without any empirical parameters \cite{Kumar2007,Kuhne2011}. Specifically, the equipotential region of the 2D PMF surface, which passes through the saddle point and encircles the minimum, corresponds to the H-bonded state, i.e. the H-bond indicator variable $B(R_{OO},\alpha_{OHO})=1$, where $\alpha_{OHO}$ the angle between $R_{OO} = \lvert \mathbf{r}_{O,a} - \mathbf{r}_{O,d} \rvert$ and the covalent OH bond vector ($\mathbf{r_H}-\mathbf{r_O}$). For all other combinations of ${R}_{OO}$ and $\alpha$, $B(R_{OO},\alpha_{OHO})=0$. 

For the purpose to study the H-bond kinetics of the individual water layers, we define the layer-specific H-bond autocorrelation function 
\begin{equation}
  C_{LL}(\tau)=\langle L_i (t_0) L_i (t_0+\tau) \rangle / \langle L_i \rangle,
\end{equation}
where $L_i(t)$ is equal to $1$ if at time $t$ a particular molecule resides in layer $i$ and $0$ otherwise. As a consequence, $C_{LL}(\tau)$ is the correlation that a H-bond, which is existing at time $t_0$ in layer $i$, is also existing at time $t_0+\tau$ within the same layer. Thus, $C_{LL}(\tau)$ eventually relaxes to zero. 

\subsection{Instantaneous Water/Vapor Interface}
\label{subsec:ssinstint}

In order to study the structure and in particular the dynamics of water as a function of distance from the surface requires a reliable definition of the water/vapor interface itself. 
However, using a surface definition based on the time-averaged density profile, the relevant spatial fluctuations in space and time of the interface location are neglected. 
Therefore, the recently proposed method to locate the instantaneous interface of Willard and Chandler is employed here \cite{Willard2010}. Instead of a time-averaged density-field, a coarse-grained but time-dependent density-field $\rho_{cg}(\mathbf{r},t)$ is constructed in terms of Gaussian functions located at the center of mass of the water molecules $\mathbf{r}_i(t)$: 
\begin{equation}
\rho_{cg}(\mathbf{r},t)=\sum_{i=1}^{N/3}(2 \pi \xi^2)^{-\frac{3}{2}}
e^{-\frac{1}{2}\left(\frac{\mathbf{r}-\mathbf{r}_i(t)}{\xi}\right)^2},
\label{cgdfield}
\end{equation}
where $\xi=2.4~\text{\AA}$ is a system-dependent coarse-graining length. 
Instantaneous interfaces can now be defined as 2-dimensional manifolds $\mathbf{s}(t)=\mathbf{r}$, for which $\rho_{cg}(\mathbf{s},t)= c$. In analogy to the Gibbs dividing surface, we have set the critical parameter $c$ to equal half of the bulk water density. 
Moreover, it is possible to deduce the proximity $a_i$ of the $i$'th water molecule from the instantaneous interface for every time-step as 
\begin{equation}
  a_i=\{ [ \mathbf{s}(t)-\mathbf{r}_i(t) ] \cdot \mathbf{n}(t) \}|_{\mathbf{s}(t)=\mathbf{s}_i(t)},
\label{proxint}
\end{equation}
where $\mathbf{s}_i(t)$ denotes the point on ${\mathbf{s}}(t)$ that is closest to $\mathbf{r}_i(t)$, while $\mathbf{n}$ is the surface normal vector at that point. Averaging over all instantaneous interfaces, the ensemble-averaged interface and the corresponding mean proximity 
\begin{equation}
\overline{a}_i = \{ [ \langle \mathbf{s} \rangle - \mathbf{r}_i ] \cdot
\langle \mathbf{n} \rangle \}|_{\langle \mathbf{s} \rangle = \langle \mathbf{s}
  \rangle _i}
\label{proxmean}
\end{equation}
is obtained. The mean density profile 
\begin{equation}
p(d)=\frac{1}{L^2}\left\langle \sum_{i=1}^{N} \delta(a_i-d) \right\rangle, 
\label{densnzeq}
\end{equation}
where $L$ is the length of the simulation cell, can be either defined in terms of $a_i$ or $\overline{a}_i$ to quantify the probability of finding a water molecule at a certain distance $d$ from the mean or instantaneous water/vapor interface. 


To characterize the orientational dependence of the water molecules on the distance to interface, we employed the joint conditional distribution:
\begin{eqnarray}
  P(u,u'|d)=\frac{1}{n(z)} &\left\langle \sum_{i=1}^{N} \delta(a_i-d)
  \delta(cos(\theta_i^{(1)})-u) \right. \nonumber \\ 
  &\times \left. \delta(cos(\theta_i^{(2)})-u') \right\rangle,
\label{jcdeq}
\end{eqnarray}
where $\theta_i^{(1)}$ is the angle between $\mathbf{n}(t)$ on $\mathbf{s}_i(t)$ and one of the two OH bond vectors $\mathbf{r}_{i,OH}^{(1)}=\mathbf{r}_{i,H}^{(1)}-\mathbf{r}_{i,O}$ of the $i$'th molecule, while $\theta_i^{(2)}$ is defined analogously for the other OH bond vectors of the same water molecule. High correlations between the surface proximity and the orientation of the water molecules appears as peaks in $P(u,u'|d)$.

\section{Computational Details}

The AIMD simulation was performed in the canonical (NVT) ensemble at 300~K using the second-generation Car-Parrinello MD method of K\"uhne et al. as implemented in the CP2K/\textsc{Quickstep} code \cite{Kuhne2007, Kuhne2014, Quickstep}. The model of the water/vapor interface consisted of a bulk water part with 384 light water molecules in a periodic orthorhombic box of dimension 15.64$\times$15.64$\times$46.92~$\text{\AA}^3$. To that, an additional vacuum portion of 8~$\text{\AA}$ on both sides along the non-periodic $z$-direction was added, while the corresponding Poisson problem was tackled by an efficient Wavelet-based solver \cite{Genovese2007}. The settings of the density functional theory (DFT) calculations to compute the interatomic forces were identical with those of our previous AIMD studies of water \cite{Kuhne2009, Kuhne2011, PascalWater2012}. The whole system was equilibrated for 1.25~ns by classical MD using the empirical SPC/Fw interactions potential \cite{PaesaniWater2006}, followed by a DFT-based re-equilibration consisting of 50~ps, before statistics was eventually accumulated for additional 250~ps. The latter was decomposed into 12 statistically independent 20~ps long trajectories, which were separated by 1~ps each. 

The classical MD and quantum mechanical PIMD simulations were conducted using a flexible water model, which was parameterized by force-matching to accurate DFT-based reference calculations using the TPSS-D3 exchange and correlation functional \cite{SpuraFM2015, Tao2003, Grimme2010}. 
The corresponding computations were all started from a well equilibrated periodic cubic simulation box with $11^3$ water molecules each and a density of 0.997~g/cm$^3$. The water surface was then created by putting two additional empty simulation cells of equal size along the $z$-direction. The resulting system was then re-equilibrated for 1~ns within the NVT ensemble at T=298~K. Thereafter, 200 MD and 130 ACMD statistical independent 20~ps long trajectories were computed in the microcanonical (NVE) ensemble. Throughout, periodic boundary conditions were employed. The short-range interactions were truncated at 9~$\text{\AA}$, while the Ewald summation technique was used to compute the long-range electrostatic interactions. In case of the ACMD simulations, the ring-polymer contraction scheme was employed with a cutoff value of 5~$\text{\AA}$ to reduce the electrostatic potential energy and force calculations to a single Ewald sum, while all other interactions were computed using $P=32$ beads. The evolution of the ring-polymer in time was performed using a discretized time-step of 0.1~fs, whereas an integration time-step of 0.5~fs was employed for the MD and AIMD simulations, respectively.

\section{Results and Discussion}

\subsection{Density Profile of the Water/Vapor interface}

\begin{figure}
\includegraphics[width=0.5\textwidth]{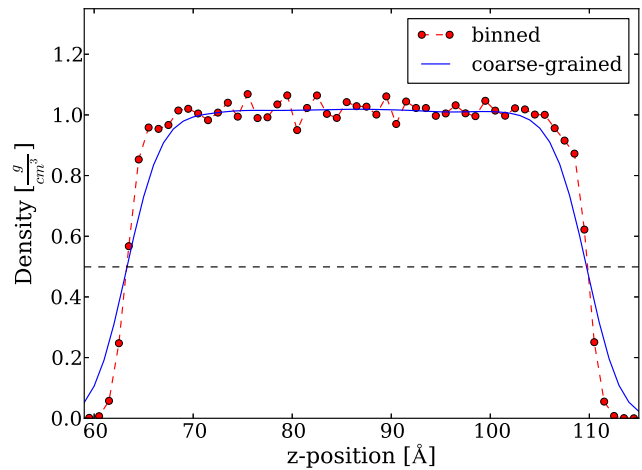}
\caption{Mean density profile of the water slab as obtained by the binned time-averaged particle density, as well as from the coarse-grained representation $\rho_{cg}(\mathbf{r},t)$. The dashed line denotes the Gibbs dividing surface, which is shown for comparison.}
\label{avdfields}
\end{figure}
In Fig.~\ref{avdfields} the time-averaged density profile of the water slab as a function of the z-coordinate is shown together with its coarse-grained counterpart $\rho_{cg}(\mathbf{r},t)$. 
\begin{figure}
\includegraphics[width=0.5\textwidth]{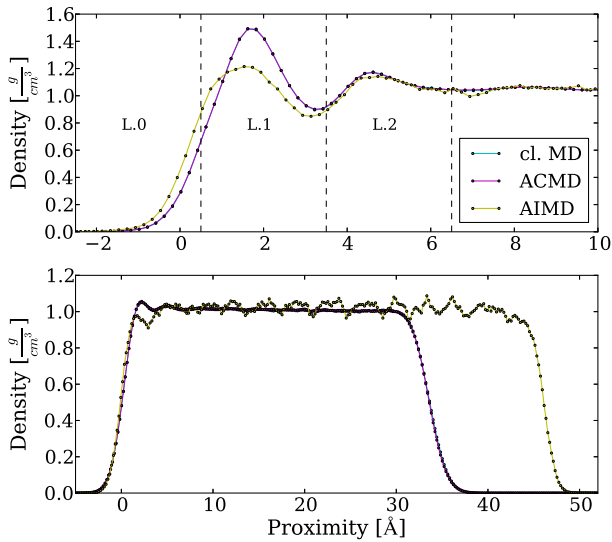}
\caption{Proximity-based density profiles in terms of the instantaneous ${a}_i$  (top panel) and mean proximity $\overline{a}_i$ (bottom panel). The instantaneous water layers (L0-L2) are indicated by vertical dashed lines.}
\label{densprof}
\end{figure}
Based on the instantaneous interface definition, the proximity-based instantaneous and mean density profiles, are depicted in Fig.~\ref{densprof}. It is apparent that the impact of NQE on the proximity-based density profiles are minor. Furthermore, using the force-matched TPSS-D3 water potential, they are more structured than those of the DFT-based AIMD simulation, which exhibits a lower first peak and a slightly larger penetration into the vapor phase. This may be attributed to the fact that compared to bulk water, the average O-O distance is increased because of relaxation effects at the water/vapor interface \cite{Stillinger1980, SurfaceRelaxation2002, kuo2004, Kuhne2011}. Interestingly, most non-polarizable interaction potentials predict the opposite, i.e. a shortening of the average O-O distance. As already described by Willard and Chandler, the density profile with respect to the instantaneous interface exhibits pronounced peaks within the interfacial region, separated by distinct minima \cite{Willard2010}. These peaks indicate that close to the instantaneous water/vapor interface, the water molecules are arranged in a layered fashion. Therefore, we will refer to the three distinct intervals of 3~$\text{\AA}$ each, as shown in the top panel of Fig.~\ref{densprof}, as instantaneous water layers (L0-L2).  However, it is important to emphasize that L0 cannot be understood as a genuine water layer, but rather as a sparse population of water molecules with a higher proximity to the vapor than to the first water layer, which is L1. In fact, despite their equal volumes, L0 consists only of about $10-20\%$ of the water molecules in L1. With this novel definition of the instantaneous water layers in the vicinity of the water/vapor interface it is now possible to compute the structure and dynamics of interfacial water for each of the various layers individually, as will be shown in the following.


\subsection{Orientation of water molecules close to the surface}

To study the orientation of the water molecules close to instantaneous water/vapor interface we employ the joint conditional distribution $P(u,u'|d)$. 
\begin{figure}
\includegraphics[width=0.5\textwidth]{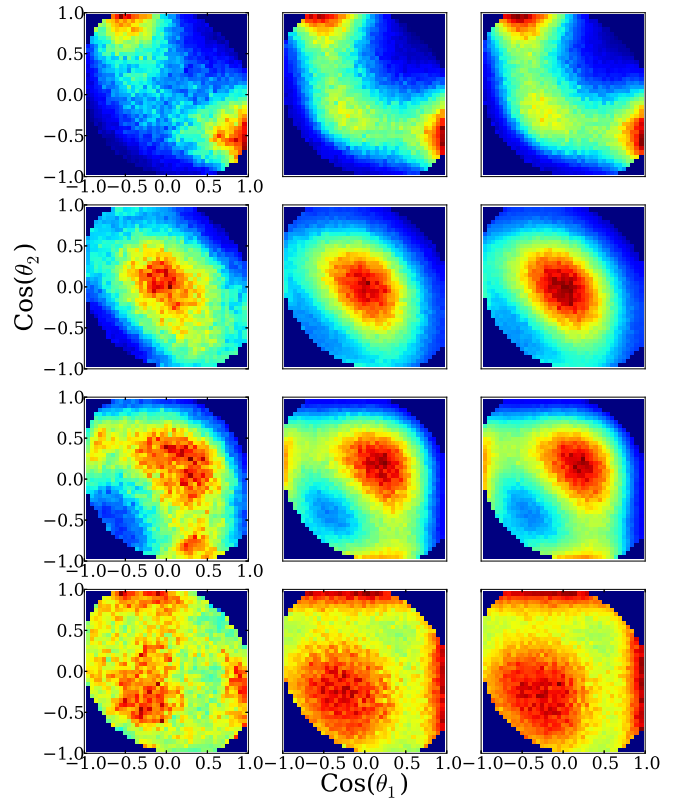}
\caption{Joint probability distribution as obtained by AIMD (left column), ACMD (middle column) and classical MD (right column) simulations. The various rows are associated with the considered proximity values of d=0.2~\AA{} to represent L0, 1.2 for L1$^{||}$ and 1.8~\AA{} for L1, as well as 4.4~\AA{} for L2.}
\label{jcdistall}
\end{figure}
The corresponding plots as a function of $\cos({\Theta})$ with respect to $\mathbf{n}(t)$, are shown in Fig.~\ref{jcdistall}. 
The various rows in Fig.~\ref{jcdistall} corresponds to four distinct proximity values between d=0~\AA{} and 5~\AA{} in order to represent the instantaneous water layers L0-L2. All layers beyond L2 do not obey any structural order and therefore correspond to bulk liquid water. As before, the influence of NQE is minor. Moreover, apart from the much enhanced statistics of our MD and PIMD simulations using the force-matched TPSS-D3 water model, no appreciable difference to the DFT-based AIMD could be detected. 

In L0 (d=0.2~\AA), $\cos(\Theta)$ is either 1 or close to -0.4, which corresponds to an angle of $0^{\circ}$ and $109^{\circ}$, respectively. From this follows that in the top-most water layer, the OH bonds are sticking out of water into the vapor phase, but are also pointing into the subjacent layer. In this L1 (d=1.8~\AA), the maxima of the joint probability distribution are antipodal to L0. Specifically, $\cos(\Theta)$ is either -1 or approximately +0.3, which equates to an angle of $180^{\circ}$ and $70^{\circ}$, respectively. As a consequence, in L1 the OH bonds are on the one hand pointing straight into the bulk phase, but on the other hand also towards L0, where the water molecules possess lone pair orbitals. While this scenario has already been predicted earlier \cite{Fan2009, Kuhne2011}, we now also find an additional region between L0 and L1 where $\cos(\Theta) \approx 0$, i.e. $\Theta \approx 90^{\circ}$. We will refer to this intermediate layer, which was first predicted by Morita and coworkers \cite{Morita2012}, where the water molecules are preferably parallel oriented with respect to the instantaneous water surface as L1$^{||}$ (d=1.2~\AA). Due to the fact that both OH bonds are perpendicular with respect to the instantaneous surface normal, both of their lone pair orbitals are pointing up- and downwards towards L0 and L1, where the associated water molecules due have their OH bonds. Moreover, in L0 and L1 an additional population at $\cos(\Theta)$ close to $\pm0.3$ is noticeable that belongs to water molecules, where both OH bonds obeys an angle of around $70^{\circ}$ and $110^{\circ}$, respectively. Put in other words, beside water molecules that are parallel with respect to the instantaneous water/vapor surface, we also find configurations that are tilted by around $\pm20^{\circ}$. The latter immediately suggest the possibility of in-plane H-bonding between the water molecules of L1$^{||}$ and the tilted configurations. The orientation of the water molecules in L2 is increasingly similar to bulk water and on the onset to become fully disordered. Nevertheless, a small orientational correlation similar to L0 is still detectable. 

\subsection{Hydrogen Bond Network Structure}

In order to investigate the structure of the H-bond network between the water molecules and to eventually devise a model of the water/vapor interface, we compute all of our ensemble-averaged H-bond network descriptors individually for all of the various instantaneous water layers. Specifically, in Table~\ref{bondtab} the average number of H-bonds, H-donors and H-acceptors per water molecule are given together with the relative fraction of inter- and intralayer H-donors, as well as the relative occurrence of ND, SD and DD H-bonds. As can be seen, the impact of NQE is again relatively small. Moreover, in the bulk water region, the MD simulations employing the TPSS-D3 interaction potential and DFT are essentially identical. However, in the vicinity of the water/vapor interface, the results between the present water model and explicit AIMD simulations differ quantitatively, which is a manifestation of the enhanced transferability of electronic structure based AIMD calculations. 
\begin{table*}
\begin{tabular}{l|c|c|c|c}
\hline
& L0 (MD/ACMD/AIMD) & L1 (MD/ACMD/AIMD) & L2 (MD/CMD/AIMD) & Bulk (MD/ACMD/AIMD) \\
\hline
H-bonds per molecule & 2.08 / 2.08 / 1.86 & 3.40 / 3.39 / 3.18 & 3.74 / 3.73 / 3.68 & 3.71 / 3.71 / 3.70 \\
H-donors per molecule & 0.96 / 0.96 / 0.84 & 1.70 / 1.70 / 1.60 & 1.84 / 1.84 / 1.82 & 1.84 / 1.84 / 1.84 \\
H-acceptors per molecule & 1.12 / 1.12 / 1.02 & 1.70 / 1.69 / 1.58 & 1.90 / 1.89 / 1.86 & 1.87 / 1.87 / 1.86 \\
Interlayer H-bonds in \% & 0.87 / 0.87 / 0.81 & 0.30 / 0.30 / 0.34 & 0.46 / 0.46 / 0.44 & --- / --- / --- \\
Intralayer H-bonds in \% & 0.13 / 0.13 / 0.19 & 0.70 / 0.70 / 0.66 & 0.55 / 0.55 / 0.57 & --- / --- / --- \\
Non-donor H-bonds in \% & 0.13 / 0.13 / 0.20 & 0.02 / 0.02 / 0.02 & 0.01 / 0.01 / 0.01 & 0.01 / 0.01 / 0.01 \\
Single-donor H-bonds in \% & 0.78 / 0.78 / 0.75 & 0.26 / 0.26 / 0.35 & 0.14 / 0.14 / 0.16 & 0.14 / 0.14 / 0.15 \\
Double-donor H-bonds in \% & 0.09 / 0.09 / 0.04 & 0.72 / 0.72 / 0.63 & 0.85 / 0.85 / 0.83 & 0.85 / 0.85 / 0.85 \\
\hline
\end{tabular}
\caption{Ensemble-averaged H-bond distribution of the individual interfacial water layers by conventional MD, quantum-mechanical ACMD and AIMD simulations.}
\label{bondtab}
\end{table*}

More importantly, we find that the water molecules in L0 form on average two H-bonds only. 
The fact that roughly 75\% of these H-bonds are SD configurations is consistent with our observation that in L0 one of the OH bonds is protruding out of the water into the vapor phase. As already mentioned, this implies that the other OH bond is pointing towards L1, which manifests itself that more than 80\% of the H-bond donors are so-called interlayer H-bonds that are connecting L0 with L1. 

In L1, the water molecules are forming on average more than three H-bonds. Of these, nearly 75\% are DD, while still approximately 25\% are SD configurations. 
Moreover, at variance to L2 and in particular to L0, about 70\% of the H-bonds are formed in-layer within L1. The large fraction of intralayer H-bonds is mainly due to the water molecules of L1$^{||}$ that are oriented parallel to the water surface, while the purpose of the remaining interlayer H-bonds is to connect the water molecules of L1 with those of L0 and in particular with L2. This implies that the water molecules in L1, which are possessing only slightly less H-bonds than in the bulk, are able to compensate the lack of potential H-bond partners owing to the presence of the vapor phase by forming additional intralayer H-bonds within L1$^{||}$. 

At variance, the H-bond network of the water molecules in L2 is rather similar to bulk water with on average nearly four H-bonds per molecule. Of these, around 85\% are DD configurations. The number of interlayer and intralayer H-bonds in L2 is nearly level, with a small preference for the latter. 



\begin{figure}
\includegraphics[width=0.5\textwidth]{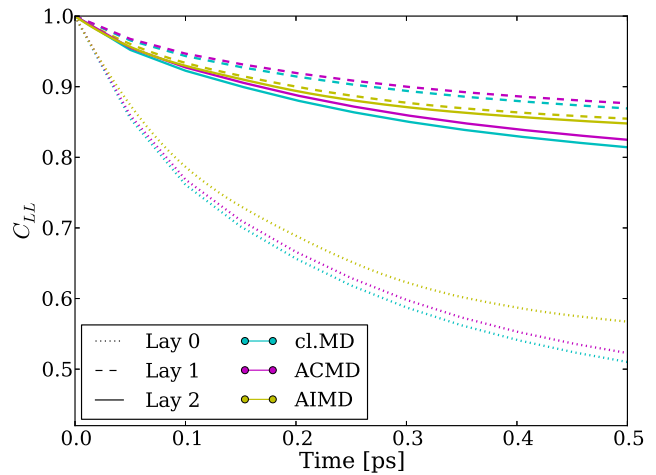}
\caption{Layer-specific H-bond autocorrelation function as obtained by conventional MD, quantum-mechanical ACMD and AIMD simulations.}
\label{inlcorr}
\end{figure}
The H-bond dynamics as determined by $C_{LL}(\tau)$ is shown in Fig.~\ref{inlcorr}. After just 0.5~ps, nearly half of the water molecules left L0, while more than $80\%$ of the molecules remained in L1 and L2, respectively. Hence, L0 is particularly mobile, whereas the H-bonds between the water molecules in L1 and especially L1$^{||}$ are particularly strong. Interestingly, employing the TPSS-D3 water model, the decay of $C_{LL}(\tau)$ for L0 and L2 is more pronounced than in our AIMD simulations, while for L1 this behavior is reversed. This implies that in the AIMD simulations, the enhanced H-bond strength of the water molecules in L1$^{||}$ is less pronounced.

\section{Conclusion}

\begin{figure}
  \includegraphics[width=0.5 \textwidth]{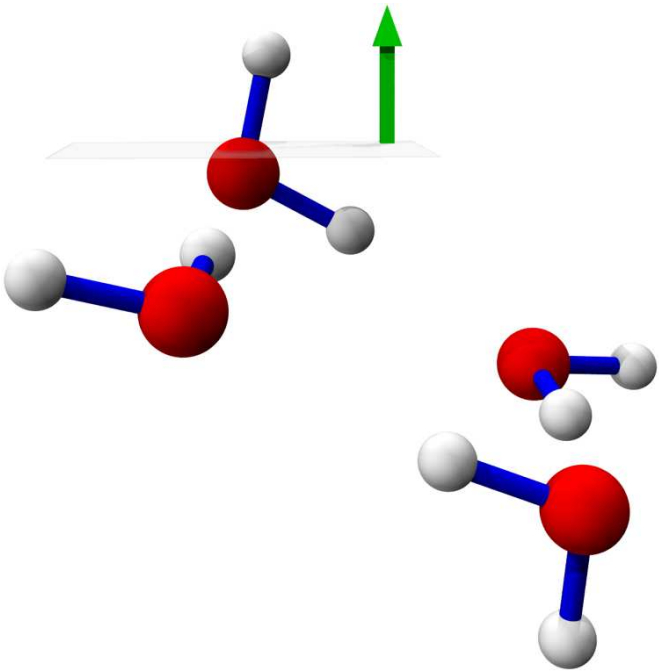}
  \caption{Schematic of the water/vapor interface from the various MD simulations. The topmost water molecule represent L0, while the lowest constitute L1. The interjacent molecules correspond to L1$^{||}$.}
  \label{InterfaceModel}
\end{figure}
Altogether, based on our simulations the following model of the water/vapor interface, which is depicted in Fig.~\ref{InterfaceModel}, is eventually emerging: 
The topmost layer of the water surface is dominated by SD water configurations, where the dangling OH bonds are preferably sticking out of the water into the vapor phase, while at the same time serving as an H-bond acceptor and donor for L1. We find that this second water layer can be further divided into the previously proposed L1 \cite{Fan2009, Kuhne2011}, where the orientation of the water molecules is inverted with respect to L0, and a novel interjacent layer denoted as L1$^{||}$. Specifically, in the former, one of the OH bonds is preferentially pointing towards the bulk-like L2, while at the same time accepting and donating H-bonds from L0 and L1$^{||}$. In this latter water layer, the molecules are oriented parallel with respect to the water/vapor interface and are able to form H-bonds with L0 and L1, as well as particularly strong intralayer H-bonds within L1$^{||}$.  
The water molecules in L2 are already structurally rather disordered and in general resembles bulk water with a relatively weak orientational correlation that is similar to L0. All of this implies that only the topmost $\sim 5$~\AA{} obeys structural oder. 
\section{Acknowledgments}
Financial support from the Graduate School of Excellence MAINZ and the IDEE project of the Carl Zeiss Foundation is kindly acknowledged. Moreover, K.K. and T.D.K. would like thank the Paderborn Center for Parallel Computing (PC$^2$) and the Gauss Center for Supercomputing (GCS) for providing computing time through the John von Neumann Institute for Computing (NIC) on the GCS share of the supercomputer JUQUEEN at the J\"ulich Supercomputing Centre (JSC). 



\bibliography{structhb}
\bibliographystyle{apsrev4-1}

\end{document}